\documentstyle[preprint,prd,aps]{revtex}

\tighten

\begin{document}

\draft

\title{The strong coupling, unification, and recent data}

\author{Paul Langacker and Nir Polonsky\thanks{Current Address:
Sektion Physik (LS Wess), University of Munich, 37 Thereisenstrasse,
D-80333 Munich, Germany.}
}

\address{Department of Physics, University of Pennsylvania,
  Philadelphia, Pennsylvania, 19104, USA}

\maketitle

\begin{center}
{UPR-0642T}
\end{center}

\begin{abstract}
The prediction of the strong coupling assuming
(supersymmetric) coupling constant unification
is reexamined. We find, using the new electroweak data,
$\alpha_{s}(M_{Z}) \approx 0.129 \pm 0.010$.
The implications of the large $\alpha_{s}$ value
are discussed.
The role played by the $Z$ beauty width is stressed.
It is also emphasized that high-energy (but not low-energy)
corrections could significantly diminish the prediction.
However, unless higher-dimension operators are assumed to be suppressed,
at present one cannot place strong constraints
on the super-heavy spectrum.
Non-leading electroweak threshold corrections are also discussed.

\end{abstract}

Assuming the minimal supersymmetric extension of the standard model
(MSSM) \cite{mssm}
between the weak and some high scale, one finds \cite{ccu}
that the extrapolated electroweak and strong couplings
approximately unify at a scale
$M_{G} \sim 3 \times 10^{16}$ GeV (the grand unification scale).
Alternatively, assuming coupling constant unification,
one can use the precisely measured
weak angle $s^{2}(M_{Z})$ and fine-structure constant
$\alpha(M_{Z})$ to predict the $Z$-pole strong coupling
$\alpha_{s}(M_{Z})$.
Model-dependent corrections are typically
of order $10\%$, i.e., comparable to
the experimental uncertainty in $\alpha_{s}(M_{Z})$,
and need to be included consistently \cite{us513}.
Below, we update and extend
our discussion of the $\alpha_{s}(M_{Z})$
prediction \cite{us513,us571,us556,us588}.
We find that for the $t$-quark pole mass
$m_{t}^{pole} \gtrsim 160$ GeV, the positive corrections
proportional to $m_{t}^{2}$ are sufficiently large that
the sum of the (Yukawa, threshold, and operator)
model-dependent corrections must cancel or be negative
for unification to hold.
Ignoring possible high-scale matching corrections,
$\tan\beta \approx 1$ and  heavy superpartners are preferred
($\tan\beta \equiv \langle H_{2}\rangle / \langle H_{1}\rangle$).
However, large negative high-scale threshold and nonrenormalizable
operator (NRO)
corrections are possible.
The former depend on the details of the grand-unified theory (GUT),
while the latter \cite{nro}
are gravitationally induced and are generic.
Below, we review our formalism and discuss our results and
their implications. We also comment on non-logarithmic
superpartner corrections, implications
of the anomalous $Z \rightarrow b\bar{b}$ width,
extended models, and on various aspects
of the large QCD coupling.
A comprehensive analysis is presented in Ref.\ \cite{diss}.

The prediction for $\alpha_{s}(M_{Z})$ reads\footnote{
Hypercharge is properly normalized, i.e., $s^{2}(M_{G}) = 3/8$.}
\begin{equation}
\alpha_{s}(M_{Z}) =
\alpha_{s}^{\mbox{\tiny OL}}(M_{Z})
+ 0.014 + H_{\alpha_{s}} + \frac{\alpha_{s}^{2}(M_{Z})}{28\pi}
+  3.1\times 10^{-7}{\mbox{ GeV}}^{-2}\left[(m_{t}^{pole})^{2}
- (m_{t_{0}}^{pole})^{2} \right]
+ \Delta_{\alpha_{s}},
\label{als1}
\end{equation}
where
\begin{equation}
\alpha_{s}^{\mbox{\tiny OL}}(M_{Z})
= \frac{ 7 \alpha(M_{Z})}{15s^{2}_{0}(M_{Z}) - 3},
\label{alsol}
\end{equation}
is the lowest order prediction,
and\footnote{We do not explicitly treat
smaller logarithmic
dependences on $m_{t}^{pole}$.
They are included in the uncertainty.
The $0.88$ factor incorporates higher-order
QCD corrections which were not included in \cite{us513}.}
\begin{equation}
s^{2}(M_{Z}) =  s^{2}_{0}(M_{Z}) -
0.88\times 10^{-7}{\mbox{ GeV}}^{-2}\left[(m_{t}^{pole})^{2}
- (m_{t_{0}}^{pole})^{2} \right],
\label{s2}
\end{equation}
where $s^{2}(M_{Z})$ is the true ($\overline{\mbox{MS}}$)
weak angle and $s^{2}_{0}$ is the value it would have for
$m_{t}^{pole} =  m_{t_{0}}^{pole}$.
The 0.014 correction is a (model-independent) two-loop gauge correction
and the function $H_{\alpha_{s}}$ is a smaller
(model-dependent) two-loop Yukawa correction.
$\alpha_{s}^{2}/28\pi$ is a finite scheme-dependent term.
The model-dependent function $\Delta_{\alpha_{s}}$
sums threshold and NRO corrections at low and high scales.
Substituting in (\ref{als1})
the ($\overline{\mbox{MS}}$) input values \cite{alpha,alpha1,fit}
$\alpha(M_{Z}) = 1/(127.9 \pm 0.1)$ and
\begin{mathletters}
\begin{equation}
s^{2}_{0}(M_{Z}) = 0.2316 \pm 0.0003,
\label{s0}
\end{equation}
\begin{equation}
m_{t_{0}}^{pole} = 160^{+11}_{-12} + 13\ln{\frac{m_{h^{0}}}{M_{Z}}}
\label{mtpole}
\end{equation}
\end{mathletters}
($m_{h^{0}}$ is SM-like light Higgs boson mass\footnote{
The authors of \cite{fit} perform a best fit to all
$W$, $Z$ and neutral current data
assuming  $60 \leq m_{h^{0}} \leq 150$ GeV with a central
value $m_{h^{0}} = M_{Z}$ for the SM-like light Higgs boson mass.
(Other possible light particle corrections are discussed separately below.)
In the (non-supersymmetric) standard model one assumes a larger
Higgs mass range $60 < m_{h^{0}} < 1000$ GeV with a central value of
300 GeV.  This leads to the prediction
$m_{t_{0}}^{pole} = 175 \pm 11^{+17}_{-19}$ GeV,
where the second uncertainty is from $m_{h^{0}}$.}),
one has (in the $\overline{\mbox{MS}}$ scheme)
\begin{equation}
\alpha_{s}(M_{Z}) -   \Delta_{\alpha_{s}}
= 0.129 \pm 0.001
+  3.1\times 10^{-7}{\mbox{ GeV}}^{-2}\left[(m_{t}^{pole})^{2}
- (160 {\mbox{ GeV}})^{2} \right]
+ H_{\alpha_{s}}.
\label{als2}
\end{equation}
The higher values of $m_{t}^{pole}$ (e.g., compared to \cite{us513})
and lower value of the weak angle implied by recent data \cite{fit}
increase the predicted $\alpha_{s}$.
An even higher central $\alpha_{s}$ value of $0.130$
would be predicted for
the value $m_{t}^{pole} = 174 \pm 16$ GeV suggested by the
CDF $t$-quark candidate events \cite{cdf}.
Two-loop Yukawa corrections are negative but are typically negligible.
They can be important if the Yukawa couplings of the $t$ and/or $b$-quark,
$h_{t}$ and $h_{b}$, respectively,
are large, i.e., for $\tan\beta \approx 1$ or
$\tan\beta \gtrsim 50$. We find \cite{us556}
\begin{equation}
-0.003 \lesssim H_{\alpha_{s}}(h_{t},\,h_{b})
 = H_{\alpha_{s}}(m_{t}^{pole}, \tan\beta) \lesssim 0.
\label{Hals}
\end{equation}
For
$h_{t} \sim \max[h_{t}(m_{t})] \sim 1.1$ (and $h_{b} \sim 0$)
one has \cite{us513}
$H_{\alpha_{s}} \sim -0.1 \times \alpha_{s}^{2} \times
h_{t}^{2} \sim -0.002$.
In general, one can substitute a one-loop semi-analytic expression
for $h_{t}^{2}$ and integrate iteratively \cite{sasha}
(a similar procedure
leads to our result for the gauge two-loop correction \cite {us513,diss}).

The coupling constant unification is shown in detail
in Fig.\ 1
for various values of $\alpha_{s}(M_{Z}) = 0.12 \pm 0.01$
and for $\Delta_{\alpha_{s}} = 0$ and $H_{\alpha_{s}} \sim -0.0005$.
In the absence of threshold corrections,
and for reasonable $m_{t}^{pole}$, coupling unification
requires $\alpha_{s}(M_{Z}) \gtrsim 0.127$.
Below, we show that typically $|\Delta_{\alpha_{s}}| \lesssim 0.01$.
Thus, we obtain from coupling constant unification,
assuming no conspiracies among different model-dependent corrections,
$\alpha_{s}(M_{Z}) \gtrsim 0.12$. This is in a good agreement
with $Z$-pole extractions of $\alpha_{s}$, but is slightly higher
than some extractions based on  deep inelastic scattering (DIS)
and quarkonium spectra. The prediction is compared with
the data in Table 1 (from \cite{table}).
The $\alpha_{s}$ measurement and the possibility
of light gluinos (that correct the $\alpha_{s}$ extrapolation
between the quarkonium and weak scales by $\sim 10\%$)
are further discussed in
Ref. \cite{qcd,shif}. We note, in passing, that light colored scalars
would correct the $\alpha_{s}$ extrapolation negligibly, i.e.,
a light scalar top would affect the extrapolation
of $\alpha_{s}$ measured at low-energy to the $Z$-pole
by less than $1\%$.
Models (in particular, NRO's)
can be constructed
with large ($\gtrsim 10 - 20\%$) and negative
GUT scale contributions
to $\Delta_{\alpha_{s}}$.
Such models would violate our no-conspiracy assumption,
but cannot be excluded. Hence,
even if supersymmetry is characterized by experiment and the
superpartner contribution to
$\Delta_{\alpha_{s}}$ (see below) is found to be positive,
coupling constant unification will not be completely ruled out
even for $\alpha_{s}(M_{Z}) \sim 0.11$.
However, one will be able to sufficiently constrain
GUT's only if the superpartner contribution is large and positive
(i.e., if NRO's with perturbative coefficients are not sufficient
to rectify the prediction).

The situation in the non-supersymmetric extension is quite different
since
({\sl a}) supersymmetry doubles the GUT sector,
({\sl b}) NRO's are typically suppressed in the non-supersymmetric case
by powers of $(M_{G}/M_{Planck}) \sim 10^{-5}$, and ({\sl c})
the corrections $\propto \alpha_{s}^{2}(M_{Z}) $ are suppressed
by a $\sim (0.07/0.13)^{2}$ factor in comparison to the MSSM
\cite{us513,diss}.
One can rectify this situation by considering large logarithms
and/or certain complicated chain-breaking scenarios with
additional particles, i.e., intermediate scales
(which, however,
could be constructed to be ${\cal O}(10^{16}$ GeV) \cite{wolf}
or ${\cal O}$(1 TeV) \cite{desh}).
The predictive power of a desert theory is lost in such a case.

Next, we discuss in greater detail the
possible
model-dependent contributions
to the ${\cal O}(10\%)$ correction function
\begin{equation}
\Delta_{\alpha_{s}}
\approx  \frac{-19\alpha_{s}^{2}}{28\pi}\ln{\frac{M_{SUSY}}{M_{Z}}}
+ {\mbox{ GUT threshold corrections $+$ NRO corrections.}}
\label{delals}
\end{equation}

The parameter $M_{SUSY}$ \cite{us513} is a weighted sum of all superpartner
and heavy Higgs boson mass logarithms
which determines the (leading-logarithm) contribution
to $\Delta_{\alpha_{s}}$ \cite{us513}
[$\Delta_{\alpha_{s}} \sim -0.003\ln(M_{SUSY}/M_{Z})$].
Specifically,
\begin{equation}
M_{SUSY} = \prod_{i}m_{i}^{-\frac{5}{38}\left[4b_{1}^{i} - \frac{96}{10}
b_{2}^{i} + \frac{56}{10}b_{3}^{i}\right]},
\label{msusy}
\end{equation}
where the index $i$ runs over all superpartner and heavy Higgs particles.
We defined the $\beta$-function coefficients
$b_{j}^{i} \equiv a(S_{i})N_{i}(j)t_{i}(j)$, where
$a(S_{i}) = \frac{1}{3}, \frac{2}{3},-\frac{11}{3}$ for a particle $i$
of spin $S_{i} = 0, \frac{1}{2}, 1$, respectively, $N_{i}(j)$
is the appropriate multiplicity, and $t_{i}(j) = 0, \frac{1}{2}, 2,3,
\frac{3}{5}(\frac{Y}{2})^{2}$ for a singlet, a particle
in the fundamental representation of $SU(N)$, an $SU(2)$ triplet,
an $SU(3)$ octet, and for $j = 1$ and a particle with hypercharge $Y$,
respectively.
Because of
mass non-degeneracies
between colored particles (whose masses are sensitive to the
gluino mass), the Higgs and Higgsino particles
(whose masses are sensitive to $\mu$),
and the scalar leptons
(whose masses are sensitive to scalar mass boundary condition),
and because of the different weights assigned to the different particles,
$M_{SUSY}$ is not simply the geometric mean of the $m_{i}$.
In particular, the negative powers in
(\ref{msusy}) imply that $M_{SUSY}$ can be (and generally is)
much smaller than the actual masses of the superpartners.
In Fig.\ 2 we calculate
$M_{SUSY}$ for more than a thousand arbitrary\footnote{We assume
universality of the soft parameters at $M_{G}$. However, similar results
for $M_{SUSY}$ hold in more general scenarios.}
MSSM's
which are consistent with the electroweak symmetry breaking,
a neutral lightest supersymmetric particle, and sparticle
masses above experimental lower bounds and below $\sim 2$ TeV
(see \cite{us594,us627}).
$M_{SUSY}$ is proportional to the Higgsino mass parameter $\mu$
\cite{carena} and is indeed lower than the actual superpartner
and Higgs boson masses.
{}From Fig.\ 2 we have the approximate upper bound
$M_{SUSY} \lesssim 250 -300$ GeV (or the lower bound
$\Delta_{\alpha_{s}}^{SUSY} \gtrsim -0.003$).

As mentioned above, $H_{\alpha_{s}}$ is large and negative for
$\tan\beta \approx 1$. Also, $M_{SUSY} \propto |\mu| \propto
\sqrt{1/[\tan^{2}\beta - 1]}$ is maximized in that region
of the parameter space ($M_{SUSY}$ is shown as a function
of $\tan\beta$ in Fig.\ 3). The proportionality factor depends on
and grows with the superpartner masses.
Thus, a heavy spectrum and $\tan\beta \sim 1$ are slightly preferred.
This observation is consistent with
$b- \tau$ Yukawa unification (which we do not require here),
which is constrained
by the interplay between the large predicted values
of $\alpha_{s}$ and the Yukawa-unification preference of moderate
$\alpha_{s}$ values \cite{diss}. (The large
QCD radiative corrections to $h_{b}$
constrain one to regions of the parameter space in which large Yukawa
coupling can partially compensate for these
corrections\footnote{Finite superpartner loops \cite{hall} modify
only the allowed large $\tan\beta$ region.}.)
In that region one has the spectacular constraint
on the Higgs boson mass $m_{h^{0}} \lesssim 100$ (110) GeV
for $m_{t}^{pole} \lesssim 160$ (175) GeV at one loop
(and a stronger bound applies at two loops) \cite{wisc,us594,diss}.

It was recently suggested that the $Z$-pole couplings should be extracted
from the data assuming the full MSSM \cite{nlo}.
This is the case if the model contains some particles
(aside from the SM-like Higgs boson)
lighter than $\sim 100 -150$ GeV.
However, assuming the heavy MSSM limit,
$SU(2)$ breaking mixing and other non-leading
effects are negligible and
our leading-logarithm formula, which
is derived using renormalization-group techniques,
is an excellent approximation.
Otherwise, light particle
(non-logarithmic) effects can be accounted for
in the same manner used to describe the quadratic $m_{t}$ dependence
\cite{us513,us588,diss}, i.e.,
by the perturbative expansion\footnote{
One could calculate the corrections
to all fitted observables, or risk a minor inconsistency
and calculate only (universal) corrections
to the input parameter ($M_{Z}$ in our case).
The latter scheme, which we follow, is sufficient for our current purposes.}
\begin{equation}
s^{2}(M_{Z}) =  s^{2}_{0}(M_{Z})
+ \frac{s^{2}_{0}(1 - s^{2}_{0})}{1 - 2s^{2}_{0}}\left[
\Delta r_{Z}^{top} + \Delta r_{Z}^{\mbox{\tiny SUSY}}\right],
\label{s22}
\end{equation}
where $\Delta r_{Z}$ \cite{sirlin} sums (universal) corrections
to the $Z$-boson mass $M_{Z}$. The leading contribution to
$\Delta r_{Z}^{top}$ is given in eq.\ (\ref{s2}) and
$\Delta r_{Z}^{\mbox{\tiny SUSY}}$ has been calculated in Ref. \cite{drees}.
In fact, it is useful to subtract from
$\Delta r_{Z}^{\mbox{\tiny SUSY}}$
leading logarithms summed by $M_{SUSY}$ and reserve
$\Delta r_{Z}^{\mbox{\tiny SUSY}}$
to denote only additional contributions
of light superpartners.
The correction function (\ref{delals}) is modified accordingly,
$\Delta_{\alpha_{s}} \rightarrow \Delta_{\alpha_{s}} -
1.16\Delta r_{Z}^{\mbox{\tiny SUSY}}$.
The different contributions to $\Delta r_{Z}^{\mbox{\tiny SUSY}}$
are correlated in a given model,
and their interplay determines its magnitude and
overall sign.
We find  \cite{progress}
that non-logarithmic
corrections typically conspire
with the $m_{t}^{2}$ term and increase the
$\alpha_{s}$ prediction, in some cases,
by a few percent.
Thus, heavy superpartners are preferred
beyond the leading order.

On a similar note,
it has been observed that if supersymmetry significantly modifies the $Z$
hadronic width (so that the $Z \rightarrow b\bar{b}$ anomaly is
accounted for) then $\alpha_{s}$ extracted from the $Z$ line shape
is diminished significantly (e.g., $0.126 \rightarrow 0.112$) \cite{fit},
and this effect was even promoted as a possible resolution
of the discrepancy between low and high-energy extractions
of $\alpha_{s}$ \cite{shif}.
Such a scenario would require
either light Higgsinos and large Yukawa couplings
or a very large
$\tan\beta$ and a light pseudo-scalar Higgs boson \cite{zbb},
i.e., $|\mu| \lesssim {\cal O}(M_{Z})$.
However, a scheme with a small $\mu$ parameter
is not favored in GUT models \cite{us627}.
{}From our discussion above it is also clear that
a solution involving light Higgsinos
(or a light pseudo-scalar)
is strongly disfavored by the $\alpha_{s}$ prediction:
\begin{enumerate}
\item
The extracted $\alpha_{s}$ line-shape value would decrease
(in agreement, however, with low-energy extractions).
\item
The predicted $\alpha_{s}$ value would
increase due to leading-logarithm [$\propto -\ln(|\mu|/M_{Z})$]
and possibly non-logarithmic threshold corrections.
\item
The central value of the fitted $m_{t}^{pole}$ [eq. (\ref{mtpole})]
would grow to 163 GeV, further increasing the $\alpha_{s}$ prediction
by 0.0003.
\end{enumerate}
Thus, the $Z \rightarrow b\bar{b}$ anomaly, if not resolved,
contains strong implications for supersymmetric models
and could even rule out the simplest and
most attractive unification scenarios.

Lastly, we consider possible high-scale contributions to the
correction function. Unlike the MSSM case, in which the
particles and their mass range are dictated by the model,
the details of the
high-scale corrections are ambiguous.
In the minimal $SU(5)$ model \cite{su5}
negative threshold corrections in (\ref{delals})
due to super-heavy color triplet Higgs
supermultiplets are strongly constrained by the non-observation
of proton decay \cite{proton},
and the GUT-scale threshold correction
contribution to $\Delta_{\alpha_{s}}$
is typically positive. (This observation, however,
need not hold in extended models.) Nevertheless, one cannot
extract strong constraints on the GUT spectrum.
Gravitationally induced operators
(suppressed by $M_{G}/M_{Planck} \sim 0.001$)
split the $M_{G}$ gauge couplings (in a correlated manner)
and correct the $\alpha_{s}$ prediction in proportion to their
effective strength, $\eta$, which is a free parameter
and can have either sign.
One has\footnote{The proportionality factor
is calculated here in the $SU(5)$ theory \cite{us513},
and its normalization is different by a factor of four than in
\cite{us513}.}
$\Delta_{\alpha_{s}}^{\mbox{\tiny NRO}} \approx 0.005\eta$.
Constraining the NRO corrections to stay perturbative
so that the calculation is consistent (higher-order terms
are negligible in this case) one has $|\eta| \lesssim 2$
($|\eta| \sim 3$ is an extreme but still acceptable choice).
Thus, NRO's with a non-negligible
and negative $\eta$ could smear
light and heavy threshold corrections. Unless $\eta \gtrsim 0$
and/or  $M_{SUSY} \ll M_{Z}$
(which could also imply positive non-logarithmic corrections),
no significant constraints
can be placed on the super-heavy spectrum at present.
On the other hand, the minimal $SU(5)$ model (where threshold
corrections are strongly constrained) would require
NRO's with $\eta < 0$
if $\alpha_{s}(M_{Z}) \lesssim 0.125$. (A similar observation was made
recently in Ref.\cite{nath}.)
Thus, unification and quantum gravity may be inseparable.

Regarding the unification scale,
corrections that increase the unification scale would typically
also increase the prediction for $\alpha_{s}(M_{Z})$ \cite{dim},
and are thus difficult to construct
[in particular, for $s^{2}(M_{Z}) \approx 0.2316$].
This is true
for contributions to $\Delta_{\alpha_{s}}$
as well as for an additional matter family
[$\alpha_{s}(M_{Z}) \rightarrow 0.132$]
or additional pairs of Higgs doublets [which lead to non-perturbative values
of $\alpha_{s}(M_{Z})$]. This is easily understood if we
write $\alpha_{s}^{\mbox{\tiny OL}}$ as a function of the unification
scale $M$ and of $\alpha(M_{Z})$ (see Fig.\ 4),
\begin{equation}
\alpha_{s}^{\mbox{\tiny OL}}(M_{Z}) = \frac{8\alpha({M_{Z}})}
{3 - 60\alpha(M_{Z})t},
\label{als3}
\end{equation}
where $t = (1/2\pi)\ln(M/M_{Z})$.
Naively substituting, e.g.,
$M = M_{\mbox{\tiny string}} \sim 5\times 10^{17}$ GeV
\cite{mstring}, one has $\alpha_{s}(M_{Z}) > 0.2$.
By carefully adjusting operator and super-heavy threshold correction
contributions to $\Delta_{\alpha_{s}}$,
one could increase $M_{G}$ by an order of magnitude
while maintaining an acceptable prediction for
$\alpha_{s}$ \cite{us571,diss}. However, in general,
to rectify the string and unification
scales (in level-one models)
one has to  compromise
the predictive power of the unification scenario \cite{string}
so that the correlation between $\alpha_{s}$ and $t$ is modified.

To conclude, we have shown that typically one expects a
large QCD coupling in supersymmetric unified models
(and even more so when considering a typical MSSM spectrum).
This constitutes an interesting signature and
has implications for, e.g., Yukawa unification, corrections
to the unification scale, and the overall sign of the correction
function $\Delta_{\alpha_{s}}$,
and is in possible conflict with low-energy  data.
However, it does not yet allow a
significant constraint on the super-heavy spectrum
because of possible gravitational corrections.
We also pointed out the interesting role that the $Z$ hadronic width
might play in supersymmetric GUT's,
and suggested a simple formula that extends
our treatment of $m_{t}^{pole}$-dependent
electroweak corrections to the supersymmetric sector.

\acknowledgments
It is a pleasure to thank S. Dimopoulos for discussions
and suggestions and to J. Erler
for discussions of precision data.
This work was supported by the US Department of Energy
Contract  No. DE-AC02-76-ERO-3071.

\begin{table}
\label{table:t1}
\caption{
Values of
$\alpha_{s}(M_{Z})$ extracted from different processes
(and extrapolated to $M_{Z}$ if relevant).
The different values are ordered according
to the energy scale of the relevant process.
}
\begin{tabular}{lc}
Bjorken sum rules & $0.122^{+0.005}_{-0.009}$ \\
$\tau \rightarrow$ hadrons (CLEO) & $ 0.114 \pm 0.003$ \\
$\tau \rightarrow$ hadrons (LEP) & $ 0.122 \pm 0.005$ \\
Deep inelastic scattering &  $ 0.112 \pm 0.005$ \\
$J/\Psi$ (lattice) & $ 0.110 \pm 0.006$ \\
$\Upsilon$ (lattice) & $ 0.115 \pm 0.002$ \\
$\Upsilon$, $J/\Psi$ (decays) & $ 0.108 \pm 0.010$ \\
$ep \rightarrow 2 + 1$ jet rate (HERA) & $0.121 \pm 0.015$ \\
$e^{+}e^{-}$ event shape (LEP) & $0.123 \pm 0.006$ \\
$Z$ line shape (LEP) & $0.126 \pm 0.005$ \\
Prediction & $0.13 \pm 0.01$ \\
\end{tabular}
\end{table}

\begin{figure}
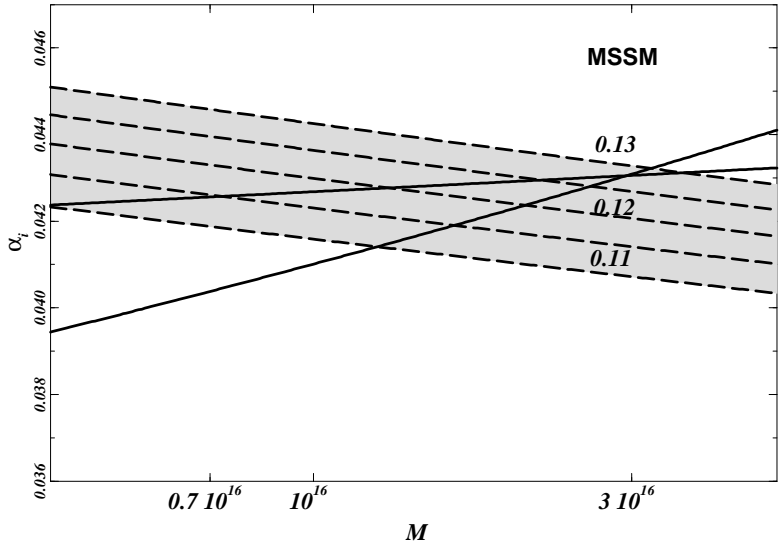

\label{fig:zoom}
\caption
{MSSM evolution of $\alpha_{1,\,2}$
(solid lines) and of $\alpha_{3}$ (dashed lines)
in the vicinity of the $\alpha_{1,\,2}$ unification point
(the scale $M$ is in GeV).
$\alpha_{s}(M_{Z})$ = 0.110, 0.115, 0.120, 0.125, 0.130;
$m_{t}^{pole} = 160$ GeV; $\tan\beta = 4$; and
$\Delta_{\alpha_{s}} = 0$.}
\end{figure}

\begin{figure}
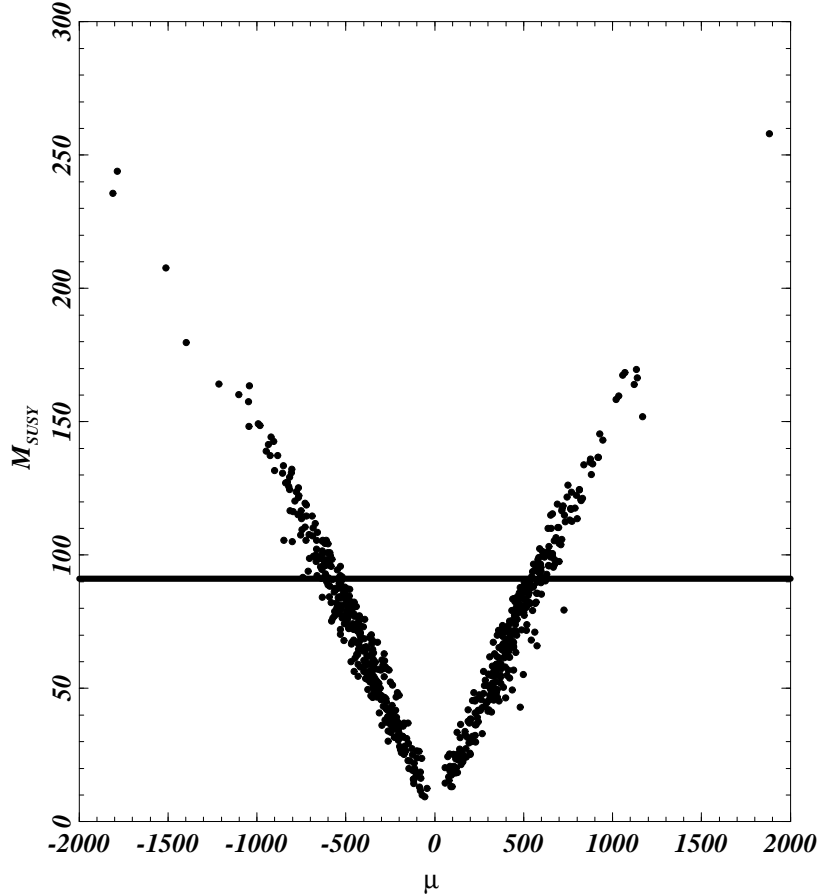

\label{fig:msusymu}
\caption
{$M_{SUSY}$ as a function of the $\mu$ parameter. The different
universal soft parameters and $\tan\beta$ are picked at random
in the allowed parameter space (see text). $m_{t}^{pole} = 160$ GeV.
$M_{SUSY} = M_{Z}$ is denoted for comparison.
(All masses are in GeV.)}
\end{figure}

\begin{figure}
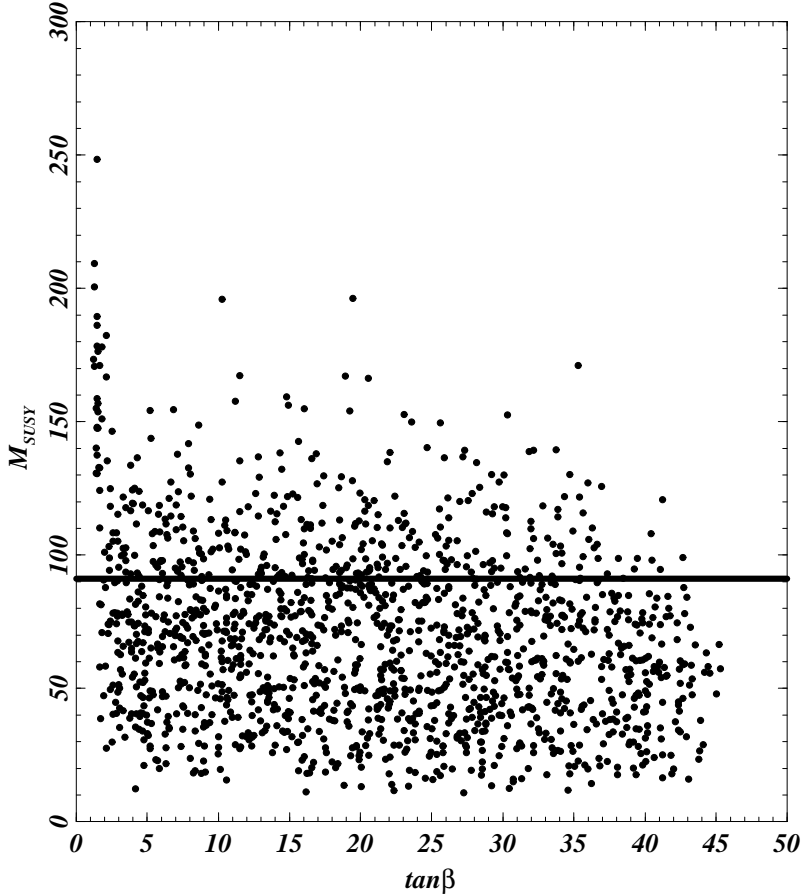

\label{fig:msusytgb}
\caption
{Same as in Fig. 2 except a function
of $\tan\beta$.}
\end{figure}

\begin{figure}
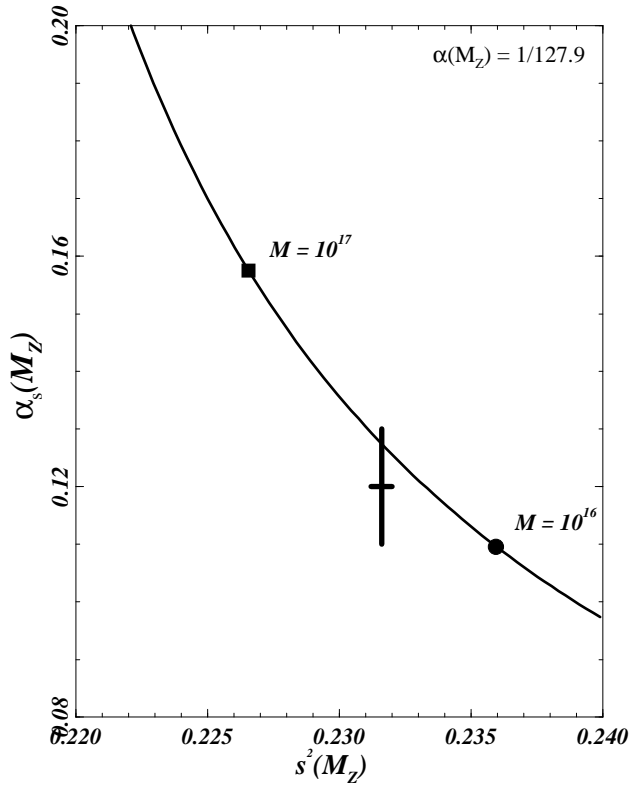

\label{fig:scale}
\caption
{
The Z-pole weak angle and strong coupling
are predicted as a function of
the unification scale $M$.
A given value of $s^2(M_{Z})$ corresponds to a fixed
choice for $M$, e.g.,
$s^{2}(M_{Z})= 0.2359$ corresponds to $M = 10^{16}$ GeV.
MSSM $\beta$-functions are assumed.
Two-loop Yukawa corrections are taken into account assuming
$m_{t}^{pole} = 160$ GeV
and $\tan\beta = 4$.
($\Delta_{\alpha_{s}} = 0$.)
$s^{2}(M_{Z}) = 0.2316 \pm 0.0003$ and $\alpha_{s}(M_{Z})= 0.12 \pm 0.01$
are indicated for comparison.
}
\end{figure}

\end{document}